\documentclass{article}

\usepackage{PRIMEarxiv}
\usepackage{amsmath} 
\usepackage{xcolor}
\usepackage[utf8]{inputenc} 
\usepackage[T1]{fontenc}    
\usepackage{hyperref}       
\usepackage{url}            
\usepackage{booktabs}       
\usepackage{amsfonts}       
\usepackage{nicefrac}       
\usepackage{microtype}      
\usepackage{subcaption}
\usepackage{lipsum}
\usepackage{fancyhdr}       
\usepackage{graphicx}       
\graphicspath{{media/}}     

\usepackage{tcolorbox}
\tcbuselibrary{listings}

\newtcolorbox{promptbox}{
  colback=black!3,
  colframe=black!40,
  fontupper=\ttfamily\small,
  boxrule=0.3pt,
  arc=2pt,
  left=4pt,right=4pt,top=4pt,bottom=4pt
}

\title{\textbf{SPAR}: Session-based Pipeline for Adaptive Retrieval on Legacy File Systems}

\author{
  Duy A. Nguyen \\
  Siebel school of Computing and Data Science  \\
  University of Illinois, Urbana-Champaign \\
  \texttt{duyan2@illinois.edu} \\
   \And
   Hai H. Do \\
  School of Communication and Information Technology \\
  Hanoi University of Science and Technology \\
  \texttt{haidh@illinois.edu} \\
  \And
  Minh Doan \\
  Bioimaging Analytics  \\
  GlaxoSmithKline, Collegeville, PA, USA \\
  \texttt{minh.x.doan@gsk.com} \\
  \And
  Minh N. Do \\
  Electrical and Computer Engineering  \\
  University of Illinois, Urbana-Champaign \\
  \texttt{minhdo@illinois.edu} \\
}

\begin{document}
\maketitle

\begin{abstract}
The ability to extract value from historical data is essential for enterprise decision-making. However, much of this information remains inaccessible within large legacy file systems that lack structured organization and semantic indexing, making retrieval and analysis inefficient and error-prone. We introduce \textbf{SPAR} (Session-based Pipeline for Adaptive Retrieval), a conceptual framework that integrates Large Language Models (LLMs) into a Retrieval-Augmented Generation (RAG) architecture specifically designed for legacy enterprise environments. Unlike conventional RAG pipelines, which require costly construction and maintenance of full-scale vector databases that mirror the entire file system, SPAR employs a lightweight two-stage process: a semantic Metadata Index is first created, after which session-specific vector databases are dynamically generated on demand. This design reduces computational overhead while improving transparency, controllability, and relevance in retrieval. We provide a theoretical complexity analysis comparing SPAR with standard LLM-based RAG pipelines, demonstrating its computational advantages. To validate the framework, we apply SPAR to a synthesized enterprise-scale file system containing a large corpus of biomedical literature, showing improvements in both retrieval effectiveness and downstream model accuracy. Finally, we discuss design trade-offs and outline open challenges for deploying SPAR across diverse enterprise settings.
\end{abstract}

\keywords{Legacy file system \and Retrieval-Augmented Generation (RAG) \and Session-based RAG}

\section{Introduction}
\label{sec:intro}

\textbf{Context.} The exponential growth of enterprise data has created an urgent need for robust systems capable of extracting, managing, and leveraging information from heterogeneous file repositories. Legacy file systems, which frequently store vast collections of historical data in unstructured or poorly structured formats, pose a persistent challenge in this regard \cite{challenge2, challenge1}. Such systems typically rely on rigid storage hierarchies and encompass diverse modalities, including unstructured documents (e.g., PDFs, DOCX), semi-structured files (e.g., JSON, CSV), and specialized formats (e.g., medical images in healthcare). The absence of semantic indexing and structured metadata further hinders their integration into modern data processing and retrieval pipelines.

\begin{figure}
    \centering
    \begin{subfigure}[t]{0.43\textwidth}
        \centering
        \includegraphics[width=\textwidth]{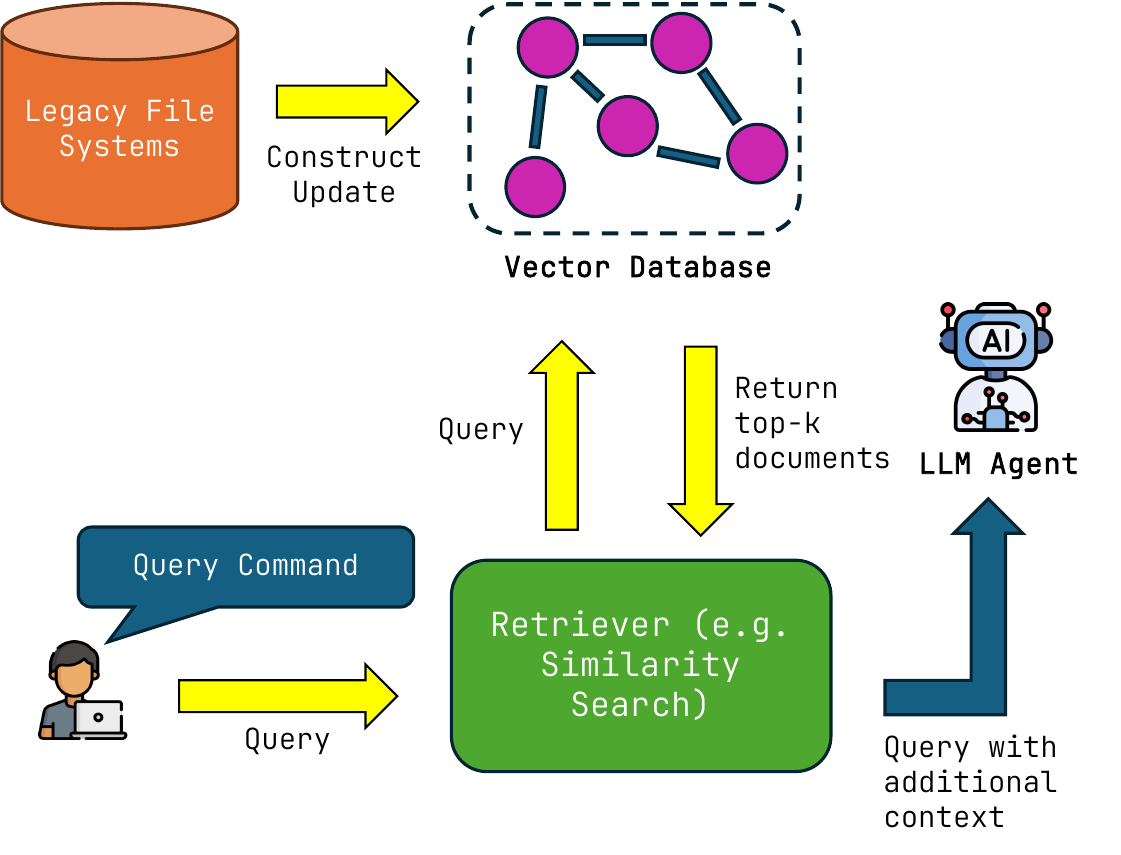}
        \caption{Ordinary LLM-based RAG}
        \label{fig:rag}
    \end{subfigure}
    \hfill
    \begin{subfigure}[t]{0.55\textwidth}
        \centering
        \includegraphics[width=\textwidth]{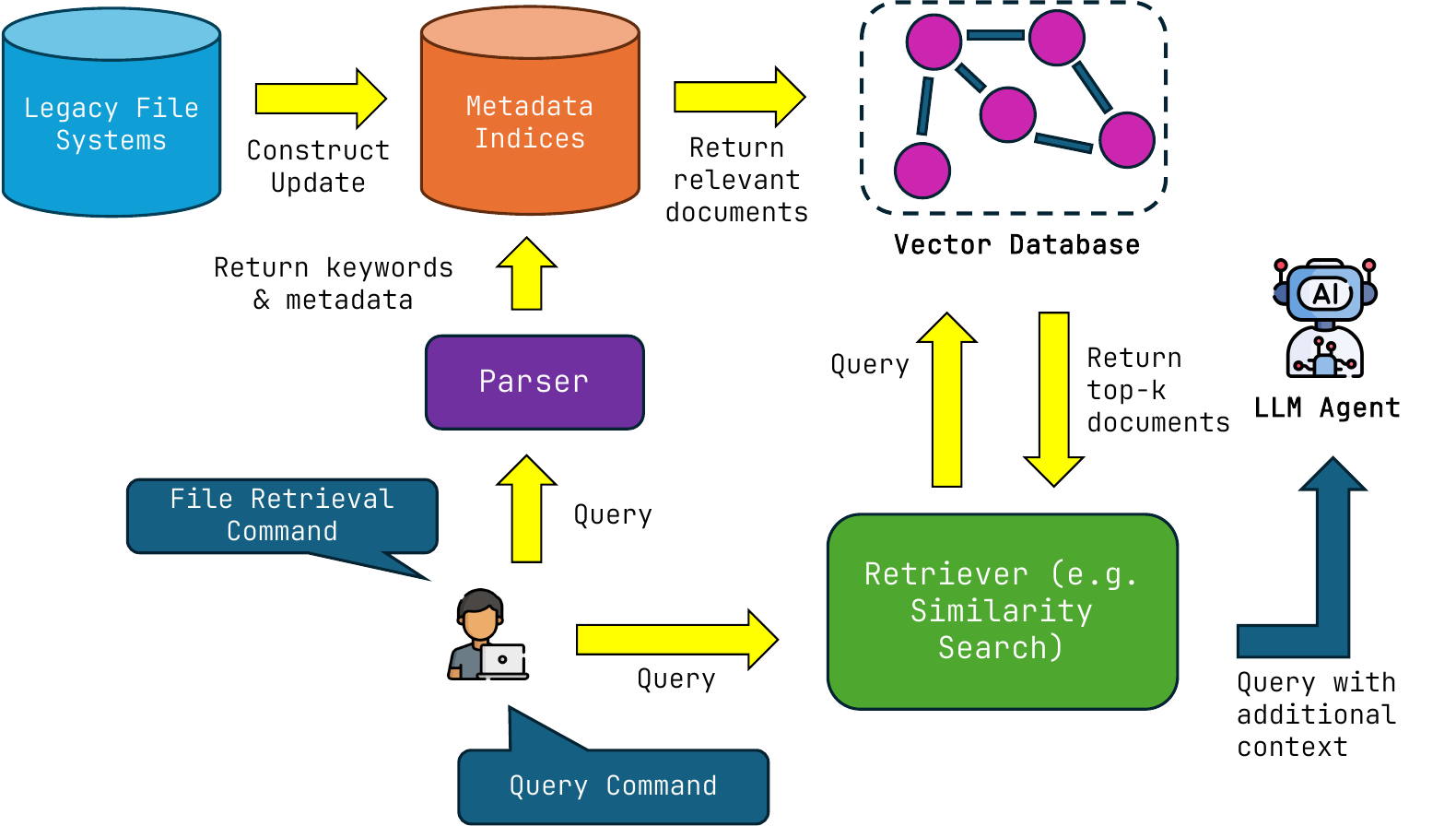}
        \caption{SPAR}
        \label{fig:spar}
    \end{subfigure}
    \caption{Overview of ordinary LLM-based RAG pipeline versus proposed SPAR pipeline.}
    \label{fig:overview}
\end{figure}

In response to these challenges, Retrieval-Augmented Generation (RAG) pipelines powered by Large Language Models (LLMs) have emerged as a promising solution \cite{rag1, rag2, rag3, rag4, rag5}. 
As shown in Figure \ref{fig:rag}, traditional RAG systems employ a two-stage process that synergistically combines retrieval and generation to augment the capabilities of LLMs, particularly for enterprise dedicated tasks.
The first stage involves constructing a comprehensive vector database, which serves as the central knowledge base for the system. This database is created by encoding data from large file systems into dense vector representations using advanced embedding models. These embeddings effectively capture the semantic relationships within the data, enabling meaningful and efficient retrieval. 
In the second stage, the system focuses on retrieval and generation. When a user submits a query, the system retrieves relevant vectors from the database through similarity search mechanisms, ranking and fetching data points that closely align with the query’s semantic meaning. These retrieved results are then passed as context to the LLM, enabling it to generate responses that are accurate, contextually grounded, and aligned with the user's needs. This retrieval step narrows the scope of information the LLM processes and tailored with enterprise context, enhancing efficiency while mitigating issues such as hallucination.

Despite its utility, this approach is not without significant limitations:
\begin{itemize}
    \itemsep0em
    \item[1.] \textbf{Resource Intensive}: Constructing and maintaining a large-scale vector database tightly mirroring an existing file system demands substantial computational and storage resources, which can be a barrier for resource-constrained systems.
    \item[2.] \textbf{Dynamic Syncing Issues}: Continuous synchronization between the file system and the vector database introduces overheads and risks of inconsistencies, especially as datasets grow in size and complexity.
    \item[3.] \textbf{Scalability Constraints}: Querying large vector datasets often results in inefficiencies, reduced accuracy, and limited user control over the relevance and specificity of retrieved outputs.
\end{itemize}
These inherent drawbacks highlight the need for more adaptive and resource-efficient solutions, motivating the exploration of innovative approaches to RAG pipeline design.

\textbf{Our method.} To address these limitations, we present SPAR (Session-based Pipeline for Adaptive Retrieval), a novel and modular RAG pipeline tailored specifically for legacy enterprise file systems (Figure \ref{fig:spar}). This approach departs from conventional RAG methods \cite{rag1, rag2, rag3} by eliminating the need for extensive, static vector databases that mirror large file systems. Instead, SPAR introduces a common strategy for constructing lightweight metadata indices that encapsulate file metadata and enterprise-defined tags. These indices are not only highly adaptable to diverse enterprise data and domains but also significantly more efficient and manageable than traditional methods.
Building on these indices, SPAR employs a session-based approach in which each query session begins with a specialized prompt that leverages the metadata to retrieve relevant files. This prompt dynamically generates a temporary vector database, customized to meet the session’s unique requirements. The resulting database acts as a flexible and persistent knowledge base for RAG operations throughout the session, with on-demand updates enabled via user prompts to ensure relevance and accuracy.

By combining the efficiency of metadata indices with the flexibility of on-the-fly vector database construction, \textbf{SPAR} addresses key limitations of traditional RAG systems:

\begin{itemize}
\item \textbf{Efficiency and Scalability}: The use of lightweight metadata indices and session-specific vector databases introduces minimal resource overhead, making SPAR adaptable to enterprise workloads of varying scale.
\item \textbf{Seamless Synchronization}: Metadata indices can be efficiently updated in real time to reflect changes in the underlying file system. Meanwhile, session-based vector databases are decoupled from persistent storage, eliminating the need for continuous synchronization.
\item \textbf{Accuracy and Usability}: Enterprise-defined tags enable precise filtering of relevant files during retrieval. The smaller scope of session-specific vector databases also enhances retrieval precision and interpretability.
\end{itemize}

Through a comprehensive theoretical analysis encompassing construction, maintenance, and synchronization complexity, we demonstrate that SPAR provides a more efficient and adaptable solution for applying RAG to legacy file systems than conventional approaches.
Beyond theory, we validate the framework on a synthesized enterprise-scale file system containing a large corpus of biomedical literature. Compared with a standard LLM-based RAG pipeline built on a global vector database, SPAR achieves higher retrieval accuracy and improved downstream model performance, underscoring its potential for practical enterprise deployment.

The remainder of the paper is structured as follows. Section~\ref{sec:prelim} reviews conventional LLM-based RAG pipelines, their extensions, and their application in enterprise settings. Section~\ref{sec:method} details the design principles and key components of the proposed SPAR framework. Section~\ref{sec:comparison} provides a complexity-based comparison between SPAR and traditional RAG pipelines. Section~\ref{sec:application} demonstrates SPAR in enterprise use cases, while Section~\ref{sec:discussion} discusses design trade-offs, deployment considerations, and open research challenges.
\section{Preliminary: LLM-based RAG pipelines}
\label{sec:prelim}

The ordinary RAG framework~\cite{rag_original, survey_rag1, rag_survey} follows a ``retrieve-then-generate'' paradigm. Enterprise data from diverse sources---such as financial reports, customer support logs, technical documentation, and other domain-specific repositories---is first ingested and preprocessed. To meet the input constraints of LLMs, the data is segmented into smaller chunks and transformed into dense vector representations using embedding models such as BERT~\cite{bert} or Sentence Transformers~\cite{t5, llama}. These vectors, which capture semantic relationships, are stored in vector databases like FAISS or Pinecone~\cite{faiss}, optimized for large-scale enterprise workloads.
When a user submits a query---for example, to generate a report, analyze business trends, or draft a customer response---the system encodes the query into a vector representation. This query vector is matched against stored embeddings to retrieve the most relevant information. The retrieved content is then passed to the generative stage, where the LLM synthesizes a response that integrates enterprise-specific evidence with its inherent generative capabilities. In this way, the final output is not only coherent and fluent but also grounded in the enterprise’s operational context.

RAG is frequently highlighted for its ability to mitigate hallucinations in LLMs by grounding them in external evidence. Within enterprise contexts, this grounding capability opens new opportunities and challenges alike~\cite{rag_enterprise1, rag_enterprise2, rag_enterprise3, rag_enterprise4}. 
On the one hand, RAG pipelines enable enterprises to incorporate diverse and unstructured data---ranging from documents and spreadsheets to knowledge repositories---into LLM-driven workflows. This integration allows organizations to transform raw data into actionable insights, producing outputs that are both operationally meaningful and tailored to enterprise-specific needs. Moreover, RAG-based systems can dynamically incorporate updates to the knowledge base, ensuring responsiveness to evolving contexts. For instance, new records from ongoing projects or real-time customer interactions can be integrated into the retrieval process without the need to retrain the underlying model.

On the other hand, ordinary RAG pipelines encounter several limitations when deployed at enterprise scale. First, the construction and maintenance of large vector databases is computationally expensive, particularly for heterogeneous file systems spanning decades of historical data. Second, exclusive reliance on similarity-based retrieval introduces usability concerns: users often lack control over the selection of retrieved content and must carefully craft queries to obtain relevant results. This misalignment between query formulations and document embeddings can lead to irrelevant or noisy retrievals. Third, as the volume of enterprise data continues to grow, system scalability is strained, reducing retrieval efficiency and degrading output precision. Consider, for example, a financial enterprise maintaining decades of annual and quarterly reports in mixed formats. When an analyst queries the system for \emph{``summarize revenue growth for Q2 2019''}, a conventional RAG pipeline may return documents discussing \emph{``quarterly revenue trends''} or \emph{``Q2 performance''} but from different years. Because such pipelines typically lack metadata-aware filtering for temporal constraints, the analyst must manually refine queries or sift through irrelevant results, leading to inefficiency and possible misinterpretation. This example highlights how seemingly minor gaps in retrieval control can have significant consequences in high-stakes enterprise settings.


Extensions to the basic RAG framework have been developed with enterprise use cases in mind, aiming to address specific shortcomings of ordinary systems. 
To enhance retrieval precision in complex environments, researchers have explored fine-grained indexing~\cite{index_1, index_2, index_3} (e.g., FKE, which retrieves sentence-level knowledge via chain-of-thought prompting~\cite{index_3}), query rewriting strategies~\cite{query_1, query_3, query_4} (e.g., Rewrite--Retrieve--Read~\cite{query_4}), and metadata-driven reranking~\cite{query_2}. 
Further improvements include hybrid retrieval systems that combine dense and sparse representations (e.g., Blended RAG~\cite{hybrid_1}), as well as diagnostic and benchmarking tools such as RAGChecker and RankArena~\cite{eval_1, eval_2} for evaluating and refining pipeline performance. 
Despite these advances, such methods typically presuppose a static, centralized vector database that mirrors the underlying file system. This assumption creates critical barriers in enterprise contexts: while indexing, query rewriting, and reranking can improve precision, they do not address the heavy computational cost of constructing and continuously synchronizing a global vector database with dynamic and evolving file systems. 

To meet domain-specific needs, modular RAG variants have also been explored in areas such as legal document analysis~\cite{legal_1}, customer support automation~\cite{customer_1}, and financial forecasting~\cite{finance_1}. 
Although such systems can deliver tailored functionality once the data has been ingested, they inherit the same scalability and maintenance bottlenecks, limiting their applicability to heterogeneous enterprise repositories that often lack structured metadata and span multiple formats accumulated over decades. 
Consequently, despite these extensions, existing RAG pipelines remain difficult to operationalize in real-world legacy environments. 

In contrast, \textbf{SPAR} overcomes these barriers by decoupling retrieval from large-scale persistent vector stores. Its use of lightweight metadata indices and session-based, on-demand vector databases directly addresses the synchronization, efficiency, and scalability issues that current extensions leave unresolved. 
The next section introduces the design principles and key components of SPAR, showing how this framework provides a practical and adaptive foundation for enterprise-scale retrieval and generation.


\section{SPAR - Session-based Pipeline for Adaptive Retrieval}
\label{sec:method}
\subsection{Overview}
\begin{figure}
    \centering
    \includegraphics[width=0.85\linewidth]{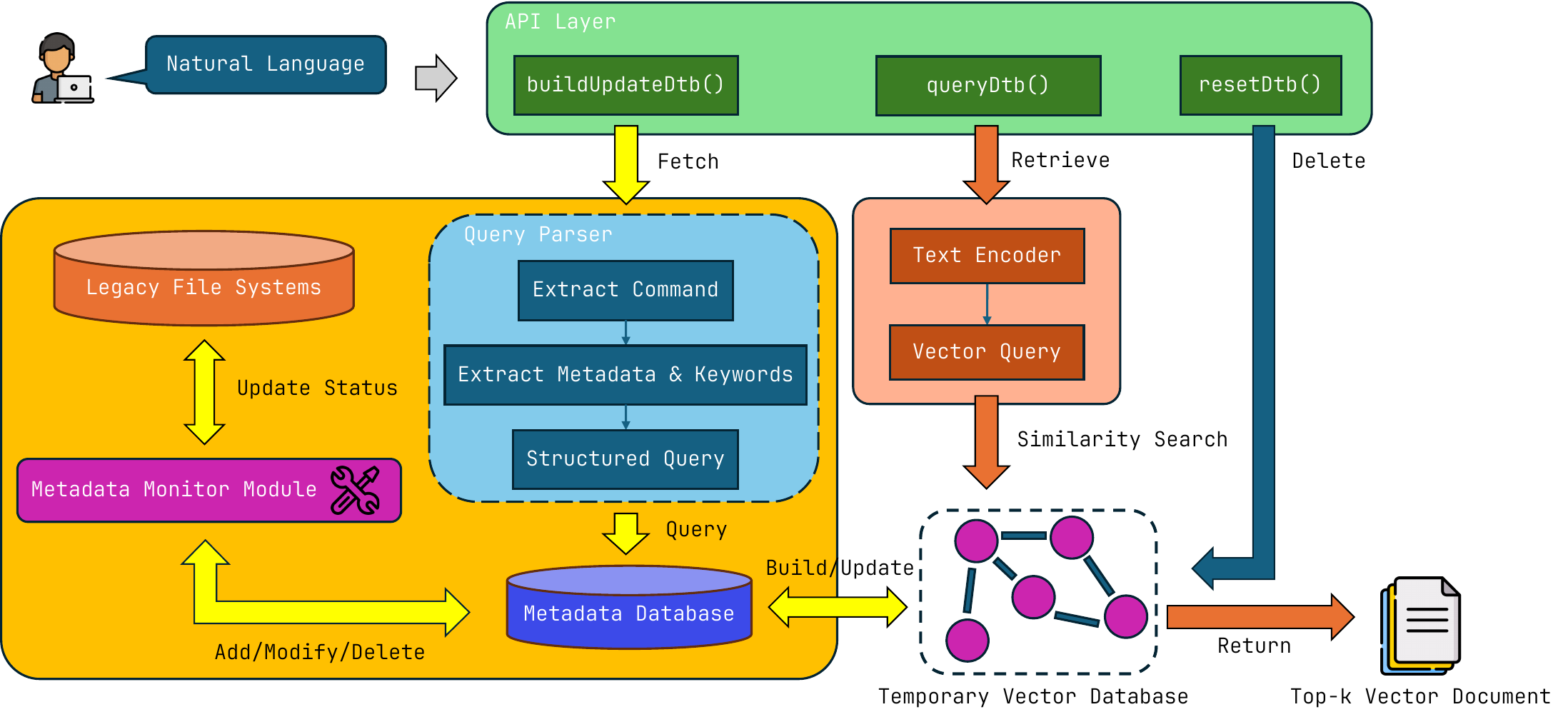}
    \caption{Detailed components of SPAR pipeline.}
    \label{fig:spar_detail}
\end{figure}

We propose \textbf{SPAR} (Session-based Pipeline for Adaptive Retrieval), a modular RAG framework designed for efficient retrieval and generation over legacy enterprise file systems (Figure~\ref{fig:spar_detail}). Unlike conventional RAG architectures that rely on static, large-scale vector databases mirroring entire file systems, SPAR adopts a flexible and lightweight strategy.  

At its core is a metadata-driven indexing scheme that encodes essential file attributes and enterprise-defined tags, yielding an enterprise-oriented \textbf{Metadata Index}. This index provides a scalable, semantically enriched representation of enterprise data and serves as the foundation for targeted retrieval and adaptive vector construction.  

Building on this index, SPAR operates in a session-oriented manner through what we call the \textbf{Session-based RAG Process}. Each user query initiates a session that begins with metadata-aware filtering to identify relevant files. From this subset, a temporary vector database is generated on demand, serving as a session-specific knowledge base. This database supports RAG operations throughout the session and can be interactively updated, enabling more precise and contextually relevant retrieval.  

Together, these two components reduce computational overhead, enhance retrieval controllability, and ensure tighter alignment with enterprise data semantics. The following subsections detail each stage in turn: first, the construction of the Metadata Index, and second, the session-based RAG process that enables adaptive retrieval and generation.

\subsection{Metadata Index construction}
\label{sec:metadata}
\begin{figure}
    \centering
    \includegraphics[width=0.9\linewidth]{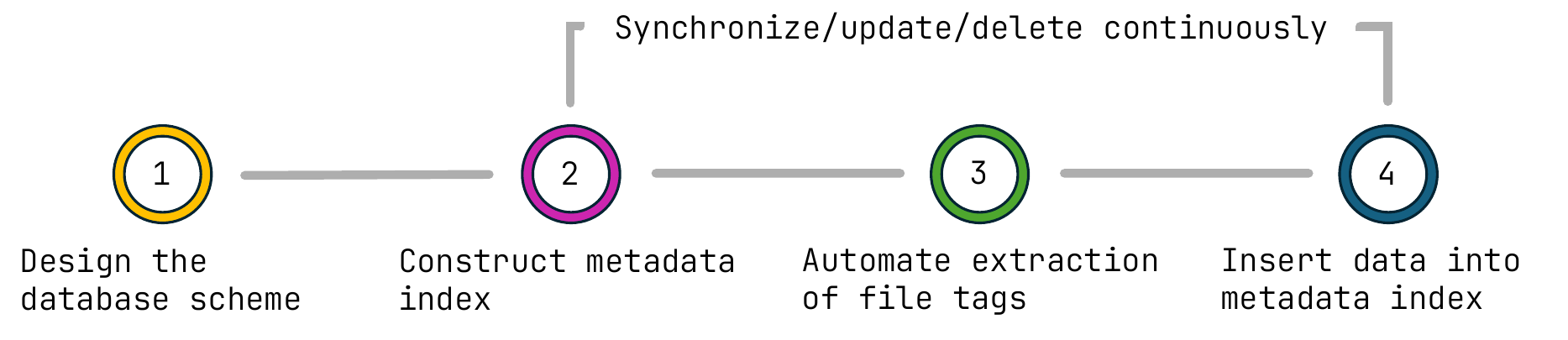}
    \caption{Conceptual steps to build and maintain a Metadata Index.}
    \label{fig:metadata_building}
\end{figure}

Instead of constructing a global vector database, we introduce a \textbf{Metadata Index}, which captures high-level descriptors of enterprise data while being more efficient and easier to update. Figure~\ref{fig:metadata_building} illustrates the conceptual steps. At the heart of the index is a hierarchical system of enterprise-defined tags and structured metadata that together enable efficient, controllable, and semantically meaningful retrieval.

\textbf{Scheme Design.}
The Metadata Index can be modeled as a relational database with two core tables: \texttt{Files} and \texttt{Tags} (Table~\ref{tab:scheme}). Each file in the legacy system is represented in \texttt{Files} by its path, metadata, and associated tags. The \texttt{Tags} table encodes both enterprise-defined and automatically extracted tags, organized into a hierarchy of parent and child concepts. This structure provides a coarse but interpretable descriptor of the file corpus, enabling fast categorization and retrieval.
\begin{table}[th]
    \centering
    \caption{Design scheme of the Metadata Index as a relational database, consisting of two tables -- \texttt{Files} and \texttt{Tags}.}
    \label{tab:scheme}
    \begin{minipage}{0.49\textwidth}
        \resizebox{\textwidth}{!}{
        \begin{tabular}{@{}ll@{}}
        \toprule
        \multicolumn{2}{c}{\texttt{Files}}                                \\ \midrule
        \texttt{File\_ID}             & Unique ID in Metadata Index             \\
        \texttt{File\_Path}           & Actual path in the system               \\
        \texttt{Tag\_ID}         & Enterprise-defined or LLM assigned tags \\
        \texttt{Metadata}         & Other metadata specific for the file \\ \bottomrule
        \end{tabular}
        }
    \end{minipage}
    \hfill
    \begin{minipage}{0.48\textwidth}
        \resizebox{\textwidth}{!}{
        \begin{tabular}{@{}ll@{}}
        \toprule
        \multicolumn{2}{c}{\texttt{Tags}}                                \\ \midrule
        \texttt{Tag\_ID}             & Unique ID in Metadata Index             \\
        \texttt{Tag\_Value}           & The tag content               \\
        \texttt{Parent\_Tag\_ID}   & IDs of parent if exist                   \\
        \texttt{Children\_Tag\_ID}  & IDs of children if exist                     \\ \bottomrule
        \end{tabular}
        }
    \end{minipage}
\end{table}

\textbf{File Tag Assignment and Hierarchy Contruction.}
\begin{figure}
    \centering
    \includegraphics[width=0.9\linewidth]{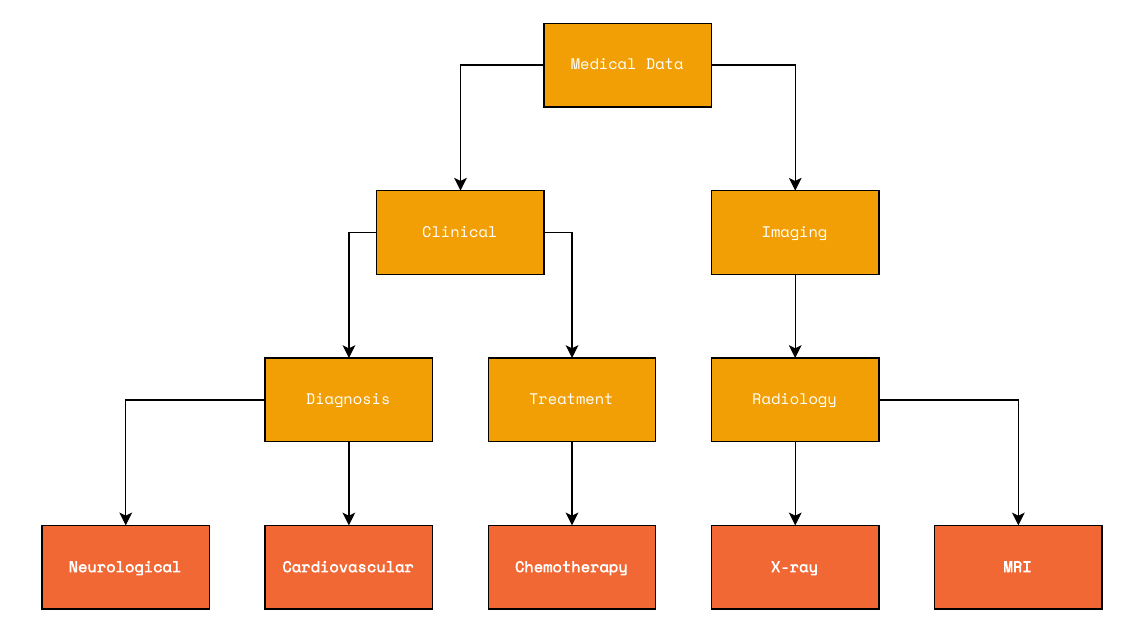}
    \caption{Example of a \textbf{Hierarchical File Tag} in a medical application. Leaf tags (orange) are defined by system owners or extracted with LLMs, while intermediate tags are generated to group related concepts. Multi-level queries allow broader or narrower retrieval based on user intent.}
    \label{fig:tag_hierarchy}
\end{figure}

\textit{Enterprise-defined flat tag set.}  
Tags constitute the primary descriptors in the proposed Metadata Index, serving to condense and represent files at a coarse level. We assume the availability of a flat set of enterprise-defined tags, corresponding to the finest level of granularity and later forming the leaf nodes in the tag hierarchy. These tags are typically specified by system owners to reflect the enterprise’s internal structure, domain logic, and organizational conventions. By design, this flat tag set encodes institutional knowledge about how information is classified and organized, ensuring that file categorization aligns with the enterprise’s operational context.  

\textit{File-tag assignment.}  
Once the tag vocabulary is established, each file in the legacy system must be mapped to one or more tags that best describe its content. This assignment can be performed manually, semi-automatically, or fully automatically using LLM-assisted tools~\cite{llm_tag1, llm_tag2, llm_tag3}. Manual assignment offers high accuracy but is costly at scale, whereas automated assignment is efficient but may introduce inconsistencies or noise. A pragmatic strategy is to combine both: LLMs provide initial assignments, while human reviewers validate a portion of them, particularly for critical datasets. This one-time effort for existing repositories is still significantly more tractable than constructing and maintaining a global vector database mirroring the entire file system (see Section~\ref{sec:comparison}). For newly ingested files, assignment can be integrated into the data pipeline as part of regular workflows, ensuring the index remains up to date. 

\textit{Additional hierarchical taxonomy.}  
To avoid flat tag sets becoming unmanageable or overlapping, we propose constructing an enterprise-specific hierarchical taxonomy. Figure~\ref{fig:tag_hierarchy} illustrates such a structure in a healthcare setting: raw tags act as leaf nodes, while intermediate nodes group related tags into broader categories based on domain logic or statistical co-occurrence. Intermediate nodes can be defined manually by domain experts or automatically induced by LLMs, which cluster semantically related leaf tags into higher-order groups. This taxonomy enables multi-level queries, where selecting an intermediate node implicitly includes all its descendant tags, allowing users to broaden or narrow retrieval scope depending on their intent. It also provides extensibility: new tags can be added over time and integrated into existing categories, maintaining structural coherence as the dataset evolves.  

Taken together, enterprise-defined tags and hierarchical organization form a principled foundation for scalable retrieval over large and heterogeneous datasets, without relying on full-text semantic embeddings. The hierarchy preserves structural coherence by reflecting enterprise-specific categorization, while supporting expressive and interpretable queries at varying levels of granularity. This design ensures transparent alignment between user intent and retrieval results, while remaining extensible as enterprise data grows.

\textbf{Metadata Characterization and Structuring.}  
Beyond tags, enterprise files often include rich metadata (e.g., modality, acquisition date, or institution for medical images; department or document type for clinical text). These attributes are consistent across workflows and should be treated as first-class retrieval constraints rather than post-hoc filters. In SPAR, metadata constraints are applied jointly with tag-based filtering, allowing users to combine conditions such as time range, modality, or department with semantic tags. This narrows the search space before embeddings are created, improving both efficiency and interpretability.

\textbf{Comparison to Traditional Approaches.}
In conventional RAG pipelines, retrieval operates over a centralized vector database, with metadata applied only after semantic similarity search. This post-hoc filtering unnecessarily enlarges the candidate set and dilutes metadata precision when embedded into dense vectors. Such designs are especially problematic for periodic or task-specific enterprise queries, where only a small, well-defined subset of data is relevant.  

SPAR eliminates the global index and instead builds a lightweight metadata index defined explicitly by tag and metadata constraints. This enables first-class filter of high-level and semantically rich information, allowing efficient retrieval of relevant subset, reduces computational overhead, and keeps enterprise-defined structure central to the process. The result is a faster, more precise, and more interpretable retrieval pipeline tailored to enterprise knowledge work. The next subsection details how this Metadata Index is leveraged during the \textbf{Session-based RAG Process}, where session-specific vector databases are dynamically constructed to support adaptive retrieval and generation.

\subsection{Session-based Retrieval RAG}
\label{sec:session_rag}


\subsubsection{Workspaces: Contextualized Environments for Targeted Retrieval}
\begin{figure}
    \centering
    \includegraphics[width=0.9\linewidth]{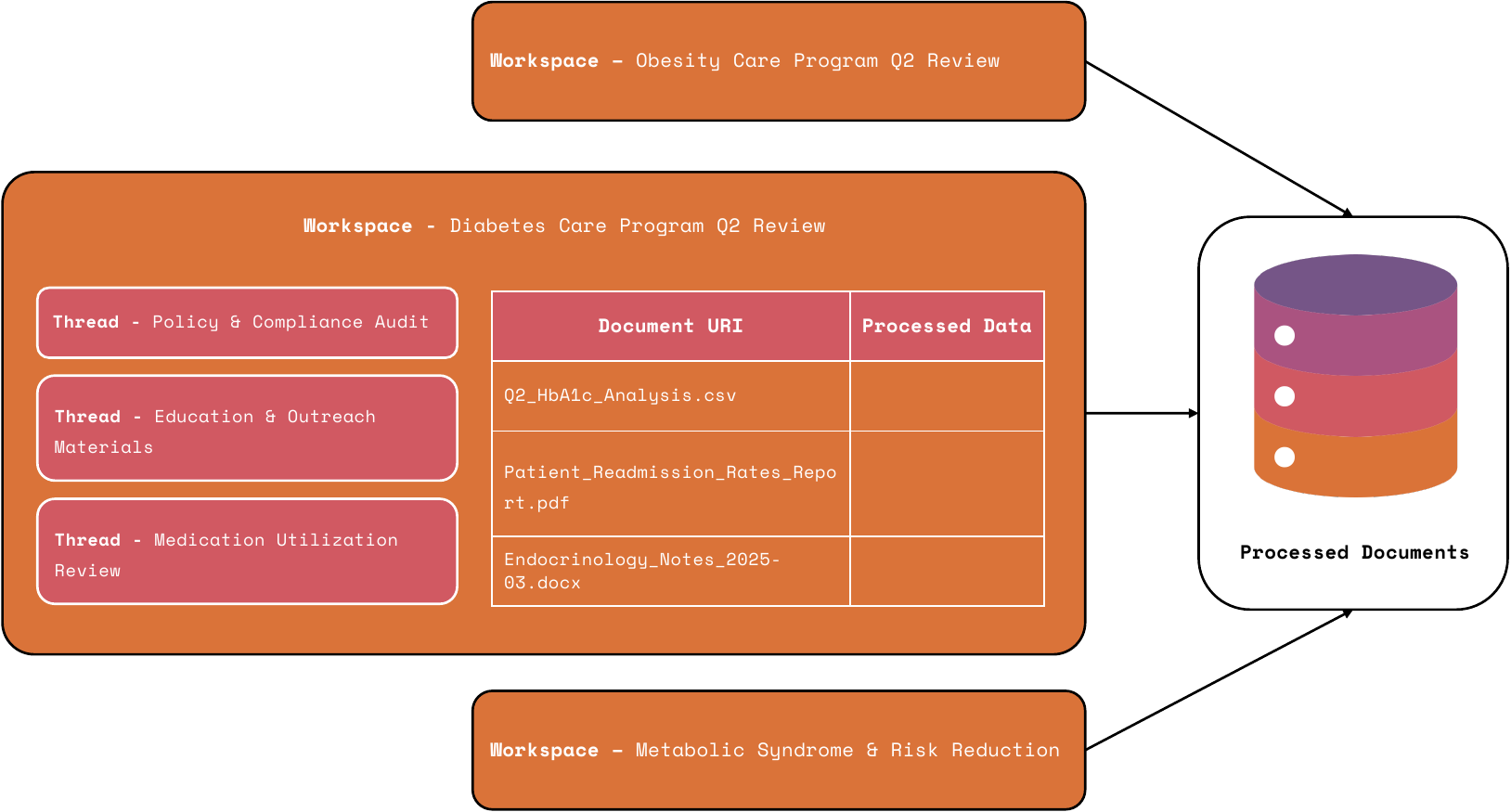}
    \caption{Example of \textbf{Session-based Retrieval} in a medical application. Each workspace corresponds to an active task with its own temporary vector database, while processed files can be cached and reused across workspaces to reduce the overhead of workspace creation.}
    \label{fig:session_illustration}
\end{figure}

To operationalize metadata- and tag-driven filtering, we introduce the concept of \textit{Workspaces}. A workspace defines a semantically scoped, time-bounded retrieval environment aligned with a specific project, reporting task, or analytic thread. Each workspace is instantiated by filtering the global corpus according to metadata and tag constraints, ensuring that only files relevant to the task are included. Figure~\ref{fig:session_illustration} illustrates this design in a healthcare application: a workspace focused on \textit{diabetes} maintains a different document pool—and thus a distinct temporary vector database—than one investigating \textit{obesity} or \textit{metabolic syndrome}. Multiple discussion threads can coexist within a workspace, all sharing its scoped database while maintaining thematic consistency.  

This design eliminates the need for a centralized global index and instead constructs lightweight, task-specific vector databases that exist only for the duration of the session. Retrieval is performed solely within this filtered subset, enabling faster and more precise query resolution. Because the candidate pool is tightly scoped from the outset, embeddings and normalized representations can be reused across queries within the same workspace, improving computational efficiency and reducing latency. Workspaces therefore provide not only targeted retrieval but also a natural structure for organizing enterprise knowledge around ongoing tasks.

\subsubsection{Main functions within workspaces}

Unlike ordinary RAG pipelines, which focus primarily on retrieval and query resolution, SPAR’s workspace framework extends the functionality beyond simple querying. In particular, it introduces two additional core operations: the creation and updating of temporary vector databases for active sessions, and their removal once the session concludes.

\begin{figure}
    \centering
    \includegraphics[width=0.8\linewidth]{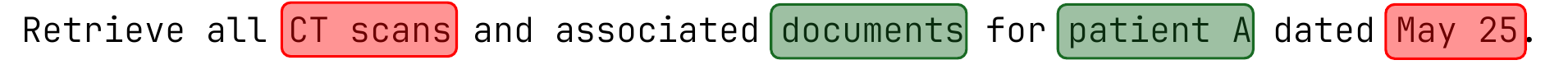}
    \caption{Example of a file retrieval prompt. \textcolor{red}{Red} text corresponds to metadata, while \textcolor{green}{green} text indicates extracted keywords.}
    \label{fig:file_prompt}
\end{figure}

\textbf{Files Retrieval and Updates.}
We introduce a dedicated command for file retrieval (\texttt{buildUpdateDtb()}, see Figure~\ref{fig:spar_detail}), which builds or updates the workspace’s temporary vector database with files relevant to the user’s query. Through this command, the user specifies the type of documents to retrieve. From the prompt, the system automatically derives filtering conditions by extracting keywords and metadata. An example prompt is shown in Figure~\ref{fig:file_prompt}, where \textcolor{red}{red} highlights illustrate extracted metadata and \textcolor{green}{green} highlights show keywords used for filtering.

To perform this extraction, a language tool ~\cite{keyword_1, keyword_4, keyword_2, keyword_3} is applied to identify salient keywords and metadata terms. Extracted keywords are embedded and aligned with the existing tag hierarchy to find semantically related tags. To avoid redundancy, a hierarchy-aware pruning step ensures that if both an ancestor and its descendant are retrieved, only the ancestor is retained. This implicitly selects all the descendant tags underlying while prevents overlapping matches and preserves broader conceptual categories--particularly useful when users express high-level intent without knowing the precise vocabulary. 

Metadata constraints (e.g., time ranges, file formats, or modality types) are mapped directly to structured filters, as they are typically explicit in the prompt and require no disambiguation.

The final filtering logic combines both dimensions: a file must match at least one tag from each selected tag group (after hierarchical expansion) and satisfy all metadata constraints to be included. This two-stage process--first by tags, then by metadata--substantially narrows the candidate pool, ensuring downstream retrieval operates only on documents aligned with user intent.

\begin{figure}
    \centering
    \includegraphics[width=0.9\linewidth]{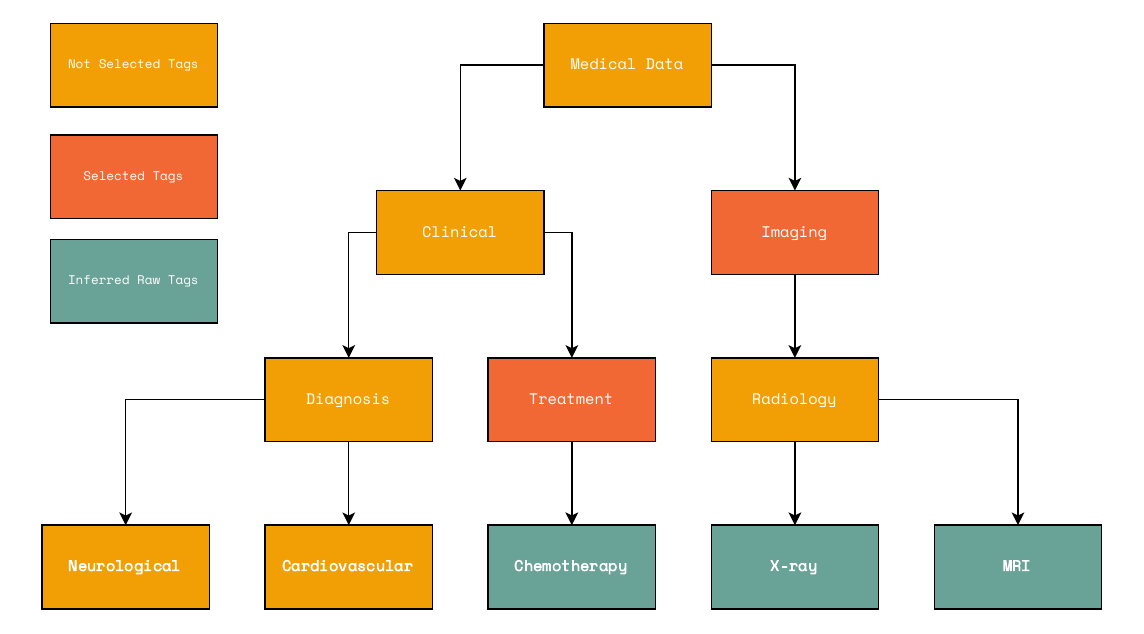}
    \caption{Example of tag selection from the hierarchy in a medical workspace.}
    \label{fig:query_interpretation}
\end{figure}

As illustrated in Figure~\ref{fig:query_interpretation}, this architecture also enhances transparency. The highlighted tag hierarchy could be returned to users and provides a clear mapping from abstract query concepts to retrieved files, helping users understand why specific documents were selected. In cases of few or no results, the structured filters can be visualized and inspected, allowing users to diagnose whether the issue lies in mismatched tags, restrictive metadata, or ambiguous phrasing—facilitating more effective query refinement.

\textbf{File processing and vector database enrichment.}
Each candidate file is first normalized into a modality-specific intermediate format (e.g., text extraction from documents, image preprocessing for scans), then embedded into a shared semantic space using a multimodal encoder ~\cite{mm_2, mm_1, mm_3}. The resulting embeddings are stored in the workspace’s temporary vector database, which serves as the basis for subsequent retrieval.

To avoid redundant computation, the system maintains a cache of normalized representations and embeddings. If a file is reused across multiple workspaces, its cached representation is retrieved directly, significantly reducing processing time and resource consumption. When a file is modified, its cached representation is automatically invalidated and refreshed during the next processing cycle, ensuring that stale embeddings are never propagated across workspaces. This mechanism supports consistent retrieval results even in dynamic file systems.

\textbf{Workspace archival and termination.}
Workspaces are designed to be temporary and task-scoped. When inactive, a workspace can either be archived or terminated. Archival removes the workspace’s vector database and intermediate embeddings while preserving metadata, filtering criteria, and access history. After a predefined expiration period, this process is triggered automatically.

Because the original file list and filtering rules are retained, a user can later reactivate the workspace and reconstruct its state without full reprocessing or manual configuration. Archived workspaces may also be shared or transferred between collaborators, enabling reproducibility and collaborative exploration without duplicating heavy preprocessing steps. This design avoids the need for continuous background updates to a global index—an issue common in traditional RAG systems, where file system changes can introduce inconsistencies and maintenance overhead. Instead, our approach shifts this burden to the workspace creation phase, which is lightweight and explicitly aligned with user context. A detailed theoretical comparison is provided in Section~\ref{sec:comparison}, with trade-offs discussed in Section~\ref{sec:discussion}.
\section{Theoretical Comparison - SPAR versus ordinary LLM-based RAG}
\label{sec:comparison}
Throughout these analyses, we follow the system of notations defined in Table \ref{tab:notation}.

\begin{table}[h]
\centering
\caption{Symbolic notations and definition for complexity analyses.}
\label{tab:notation}
\begin{tabular}{ll}
\toprule
\textbf{Symbol} & \textbf{Definition} \\
\midrule
\( N \) &
\begin{tabular}[t]{@{}l@{}}
 Total number of files provided by the enterprise.
\end{tabular} \\
\addlinespace

\( M \) &
\begin{tabular}[t]{@{}l@{}}
 Total number of predefined tags used to annotate files. \\
 The tag space is semantically meaningful and interpretable.\\
 Typically, \( M \ll N \)
\end{tabular} \\
\addlinespace

\( T_{\text{proc}} \) &
\begin{tabular}[t]{@{}l@{}}
 Average time required to process a single file. \\
 Includes format conversion and embedding of the resulting content.
\end{tabular} \\
\addlinespace

\( N_{\text{candidates}} \) &
\begin{tabular}[t]{@{}l@{}}
 Number of files returned by the metadata/tag index for a query (coarse filtering). \\
 This is the candidate set \emph{before} policy/format/quality gates and cache checks. \\
 Typically \( N_{\text{filtered}} \le N_{\text{candidates}} \); \\
 when there are no extra gates or cache reuse, \( N_{\text{candidates}} = N_{\text{filtered}} \).
\end{tabular} \\
\addlinespace

\( N_{\text{filtered}} \) &
\begin{tabular}[t]{@{}l@{}}
 Number of files actually admitted to the workspace \emph{after} policy/format/quality gates \\
 and cache checks (i.e., those we (re)process/embed and index). \\
 Typically \( N_{\text{filtered}} \ll N \), due to early-stage narrowing.
\end{tabular} \\
\bottomrule
\end{tabular}
\end{table}

\subsection{Construction and Maintenance Cost}
\label{sec:compare_construction}
\subsubsection*{Traditional RAG Pipeline}

In traditional RAG systems, the entire file corpus must be preprocessed and indexed prior to retrieval. This involves:

\begin{itemize}
    \item \textbf{Preprocessing time:} \( O(N \cdot T_{\text{proc}}) \)
    \item \textbf{Index construction:} Using an algorithm like HNSW, which requires \( O(N \log N) \) time in low-dimensional settings~\cite{hnsw}
\end{itemize}

Thus, the total construction time is:
\[
T_{\text{RAG}} = O(N \cdot T_{\text{proc}}) + O(N \log N)
\]
\subsubsection*{Proposed Approach}

SPAR performs targeted construction by leveraging enterprise tags and metadata as a first-stage filter, followed by just-in-time processing and indexing per user session (workspace).

\begin{itemize}
    \item \textbf{Offline tag index:} Built once by embedding the enterprise tag vocabulary and constructing an index (e.g., HNSW), with a one-time cost of \( O(M \log M) \).

    \item \textbf{Per-workspace vector database construction:} Each user file retrieval command triggers the following steps:
    \begin{itemize}
        \item \textbf{Tag search:} Performed via approximate nearest neighbor search over the tag index using HNSW, yielding complexity \( O(\log M) \).
        
        \item \textbf{File filtering:} 
        Via the Metadata Index (see Section \ref{sec:metadata} and Table \ref{tab:scheme}), we perform indexed lookups. The cost is
        $O\left(N_{\text{candidates}}\right).$
        
        \item \textbf{Preprocessing:} Each filtered file is processed and embedded, contributing \( O(N_{\text{filtered}} \cdot T_{\text{proc}}) \) time.

        \item \textbf{Indexing:} The resulting embeddings are added to a session-specific HNSW index, requiring \( O(N_{\text{filtered}} \log N_{\text{filtered}}) \) time.
    \end{itemize}
\end{itemize}
\noindent\textbf{Per-session construction time.}
Let \(p := N_{\text{filtered}}/N\) denote selectivity and assume indexed filtering so that \(N_{\text{candidates}}\approx N_{\text{filtered}}=pN\). Then the per-session cost decomposes as
\begin{align}
T_{\text{ours}}
&= \underbrace{O(M\log M)}_{\text{one-time tag index}} + \underbrace{O(\log M)}_{\text{tag lookup}} + \underbrace{O(N_{\text{candidates}})}_{\text{indexed filtering}} \nonumber\\
&\quad + \underbrace{O\!\left(N_{\text{filtered}}\,T_{\text{proc}}\right)}_{\text{(re)processing/embedding}} + \underbrace{O\!\left(N_{\text{filtered}}\log N_{\text{filtered}}\right)}_{\text{workspace ANN index}} \label{eq:tours-full}\\
&\approx O(M\log M)\;+\;O\!\left(N_{\text{filtered}}\left[1 + T_{\text{proc}} + \log N_{\text{filtered}}\right]\right), \tag{\(\star\)}
\end{align}
which shows that \(T_{\text{ours}}\) scales with the \emph{filtered} set size \(N_{\text{filtered}}\) rather than the global corpus size \(N\).

\noindent\textbf{Break-even (back-of-the-envelope).}
Comparing a one-time ordinary build against \(W\) SPAR sessions/workspaces, SPAR is preferable whenever
\begin{equation}
O\!\left(N\,T_{\text{proc}} + N\log N\right)
\;>\;
O(M\log M) + \sum_{w=1}^{W} O\!\left(N_{\text{filtered},w}\left[1 + T_{\text{proc}} + \log N_{\text{filtered},w}\right]\right).
\label{eq:breakeven}
\end{equation}
If sessions have similar selectivity \(p\) so that \(N_{\text{filtered},s}\approx pN\), this reduces to the rule of thumb
\(
W\,p \ll 1
\),
i.e., a few targeted sessions over narrow slices of the corpus amortize better than a full global prebuild.

\subsubsection*{Maintenance / ingestion}

For ordinary RAG, when \(k\) files arrive or change, the system pays
\[
O\!\left(k\,T_{\text{proc}}\right) \;+\; O\!\left(k\log (N+k)\right)
\]
to (re)embed and insert them into the global ANN index.
For SPAR, the Metadata Index is updated online with cost
\[
O\!\left(\mathrm{tag\_assign}(k)\right),
\]
while vector embeddings are computed just-in-time during sessions (no global vector reindex is required). This design removes the “fall-out-of-sync” failure mode that can trigger costly full reconstructions in ordinary pipelines.

\subsection{Search Time}

\textbf{Traditional RAG pipeline.}
In text-book analyses, graph-based approximate nearest-neighbor (ANN) methods such as HNSW offer (near-)logarithmic query time, e.g., \(O(\log N)\) under low effective dimensionality and fixed graph degree~\cite{hnsw}.
In modern LLM-/VLM-derived embeddings, however, the \emph{effective} dimension is high; distance concentration and sparsity dilute ANN guarantees (``curse of dimensionality''~\cite{curseofdimensionality}). In practice, maintaining recall requires increasing search-time hyperparameters—e.g., HNSW’s dynamic candidate list size \texttt{ef} (\texttt{efSearch}) and its maximum neighborhood degree \(M\)—or analogous beam/queue widths in other graph ANNS such as NSG and DiskANN; these raise per-query constant factors and can push latency beyond interactive budgets~\cite{ann_1, ann_2, ann_3, ann_4}.

\medskip
\textbf{SPAR’s selective search.}
SPAR narrows the search space \emph{before} vector lookup using the Metadata Index (Section \ref{sec:metadata}), yielding a filtered set of size \(N_{\text{filtered}}\) (Table~\ref{tab:notation}), typically \(N_{\text{filtered}}\!\ll\!N\).

Let \(T_{\text{ANN}}(N,d,\theta)\) denote the empirical cost of an ANN query as a function of database size \(N\), effective dimension \(d\), and tuning parameters \(\theta\) (e.g., \texttt{efSearch}, graph degree).
Then a single query costs
\[
T_{\text{RAG}}^{(q)} \approx T_{\text{ANN}}(N,d,\theta)
\qquad\text{vs.}\qquad
T_{\text{SPAR}}^{(q)} \approx T_{\text{ANN}}(N_{\text{filtered}},d,\theta'),
\]
with \(\theta'\) free to be more recall-oriented because \(N_{\text{filtered}}\) is much smaller.
Since \(T_{\text{ANN}}\) is (empirically) increasing in both \(N\) and \(\theta\), there exists a setting \(\theta'\!\ge\!\theta\) (increased recall-oriented knobs - e.g., \texttt{efSearch}, graph degree) such that
\[
T_{\text{ANN}}(N_{\text{filtered}},d,\theta') \;\le\; T_{\text{ANN}}(N,d,\theta),
\]
while achieving equal or higher recall.  This mitigates the high-dimensionality penalty observed in large, global indices.

\begin{itemize}
    \item \textbf{Latency-preserving regime:} Fix recall and reduce latency by replacing \(N\) with \(N_{\text{filtered}}\).
    \item \textbf{Quality-raising regime:} Keep latency flat and increase recall by raising \(\theta\!\to\!\theta'\) (e.g., larger \texttt{efSearch}) made affordable by the smaller \(N_{\text{filtered}}\).
\end{itemize}

\noindent\textbf{Amortization within a workspace.}
Each workspace builds an ANN index over its filtered set once at cost
\(O\!\big(N_{\text{filtered}}\log N_{\text{filtered}}\big)\).
For \(Q\) downstream queries in the same session, the effective per-query overhead is
\[
O\!\left(\frac{N_{\text{filtered}}\log N_{\text{filtered}}}{Q}\right),
\]
which is negligible for interactive \(Q\) (e.g., multi-turn retrieval or re-ranking passes). Filtering time itself is accounted in Section \ref{sec:compare_construction} via the Metadata Index and does not scale with \(N\) (linear scans are avoided).

By shrinking the candidate set \emph{before} vector search, SPAR avoids the scalability bottlenecks of ever-growing global indices and enables a favorable speed–quality trade-off: either faster responses at fixed quality or higher-quality results at fixed latency, with index-build costs amortized over session queries.

\subsection{Memory Usage}

The memory footprint in vector-based retrieval systems consists of two main components: the storage of the embedding vectors themselves and the additional overhead of the indexing structure (e.g., graph connections in HNSW). Let $v$ be bytes per embedding and $o$ be index overhead per vector.

\textbf{Traditional RAG pipeline.} All \( N \) embeddings are stored once in a global vector index, resulting in:
\[
\mathrm{Mem}_{\text{global}} = N \cdot (v + o)
\]

\textbf{SPAR.} Each workspace maintains a local index over a smaller filtered subset \( N_{\text{filtered}} \), but embeddings may be duplicated across multiple indices if the same file is relevant to multiple workspaces. 
If \(W\) workspaces are active, the RAM used by workspace-scoped indices is
\[
\mathrm{Mem}_{\text{SPAR}} \;\approx\; \delta \cdot \Big(\mathbb{E}[\#\text{unique files across active workspaces}]\Big)\cdot (v+o),
\]
where \(\delta\in[1,W]\) is the average duplication factor (how many workspaces, on average, index the same file).

A sufficient ``win'' condition vs.\ a global index of size \(N\) is
\[
\delta\cdot W\cdot \mathbb{E}[N_{\text{filtered}}] \;<\; N.
\]

It should be noted that the memory footprint in our approach is further optimized by the workspace archival mechanism. Inactive workspaces can be archived, which removes the associated vector index and intermediate embeddings while retaining only the necessary metadata and access history. This reduces \(\delta\) further over time and ensures no unnecessarily memory.

\section{Application in synthesized biomedical literature corpus}
\label{sec:application}
To illustrate and provide a preliminary quantitative assessment of the SPAR pipeline, we synthesize a small file system containing biomedical literature. On this corpus, we generate a limited set of multiple-choice questions to evaluate both file retrieval accuracy and model performance in answering questions with the retrieved documents, comparing against a conventional LLM-based RAG pipeline. This experiment is intended solely as a toy example and does not reflect the full application scope of SPAR, which is designed for deployment on enterprise datasets at much larger scale.

\subsection{Dataset Description}
For this experiment, we sample $1000$ full-text articles from the PMC Open Access Subset~\cite{pmc_open_access}, which contains millions of biomedical papers in raw text format released under permissive reuse licenses. Each article is linked to a PubMed ID (PMID), which we use to retrieve additional metadata, including MeSH descriptors, from the 2025 PubMed Baseline dataset~\cite{pubmed_baseline_2025}.

\subsection{Baseline Construction}
We construct both an ordinary RAG system and our SPAR pipeline to work with this toy corpus for comparison. Both systems are backed by a same LLM agent - Qwen2.5-VL 3B \cite{qwen_vl} to ensure no model bias is introduced.

\textbf{Ordinary RAG system.}  
For the baseline, we construct a global vector database covering the entire set of $1000$ articles. Each article is split into passages using a fixed window size, and the resulting chunks are embedded with a sentence-level encoder to produce dense representations. These embeddings are indexed in Pinecone database~\cite{pinecone} and serve as the knowledge base for retrieval. At query time, the user question is embedded, nearest neighbors are retrieved from the global index, and the retrieved passages are provided as context to the LLM for answer generation. This setup mirrors the standard RAG pipeline widely used in practice, where a single, centralized vector store is maintained across the entire corpus.

\textbf{SPAR system.}
To correctly set up the SPAR pipeline, the major component needed to construct is the tag hierarchy and Metadata Index.  
  
\textit{MeSH Hierarchy Construction.}  
To substitute for enterprise-defined tags, we adopt the Medical Subject Headings (MeSH) taxonomy as the structured tag system for biomedical literature. MeSH is organized hierarchically through \texttt{TreeNumber} identifiers, which encode parent–child relationships. For example, \texttt{G07} is a parent of \texttt{G07.025}, which in turn is a parent of \texttt{G07.025.133}. Table~\ref{tab:mesh-hierarchy} illustrates this structure with the descriptor \textit{Adaptation, Physiological} (\texttt{D000222}), which appears under both \texttt{G07.025} and \texttt{G16.012.500}, and itself serves as a parent to terms such as \textit{Acclimatization} and \textit{Body Temperature Regulation}. This overlapping DAG structure naturally supports hierarchical tag expansion, enabling queries to be broadened or narrowed based on user intent.  

To operationalize this hierarchy, we traverse the MeSH XML descriptors and store them in the \texttt{Tags} table of our Metadata Index. Hierarchical relationships are derived by comparing \texttt{TreeNumber} values: if one is a prefix of another, it is treated as an ancestor. In parallel, each MeSH descriptor is embedded into the vector database to support semantic similarity during file retrieval latter. Embeddings are constructed using both the descriptor’s name and its associated metadata fields (e.g., notes, annotations), yielding richer semantic representations.  

\begin{table}[ht]
\centering
\caption{Excerpt of MeSH hierarchy illustrating the DAG structure centered on \textit{Adaptation, Physiological}.}
\label{tab:mesh-hierarchy}
\begin{tabular}{lll}
\toprule
\textbf{TreeNumber} & \textbf{MeSH ID} & \textbf{Term Name} \\
\midrule
G07                 & D010829          & Physiological Phenomena \\
G16.012             & D000220          & Adaptation, Biological \\
\textbf{G07.025}    & \textbf{D000222} & \textbf{Adaptation, Physiological} \\
\textbf{G16.012.500}& \textbf{D000222} & \textbf{Adaptation, Physiological} \\
G07.025.133         & D000064          & Acclimatization \\
G16.012.500.133     & D000064          & Acclimatization \\
G16.012.500.535     & D001833          & Body Temperature Regulation \\
\bottomrule
\end{tabular}
\end{table}

\textit{Metadata Index Finalization.}  
Finally, the curated files from the previous step are linked to their corresponding MeSH descriptors by populating the \texttt{Files} table with file paths, metadata fields, and associated MeSH IDs. This completes the Metadata Index, equipping the system to perform tag-based filtering, scoped embedding, and hierarchical navigation of biomedical content at query time. Implementation details, including specific databases and tools, are provided in Appendix~\ref{sec:appendix_implementation}.

\subsection{Experimental Setup}
To evaluate SPAR against a conventional RAG baseline, we design an experiment that jointly measures retrieval accuracy and downstream answer accuracy. The evaluation relies on a set of automatically generated multiple-choice questions, each tied to a specific article in the corpus.  

\textbf{Question design.}  
From the full set of $1000$ biomedical articles, we select a subset of $335$ target articles that collectively cover a broad range of MeSH tags, ensuring representative coverage of the corpus. The remaining $665$ articles serve as distractors, thereby introducing realistic noise into the retrieval process. For each target article, a single multiple-choice question is generated using a larger variant of the chosen LLM agent—Qwen-2.5VL 7B~\cite{qwen_vl}. Each question consists of four candidate answers, with exactly one correct option, and is grounded in the main text of the associated article. The prompting procedure used for question generation is detailed in Appendix~\ref{sec:appendix_implementation}.  

\textbf{Retrieval accuracy.}  
We measure retrieval performance by computing the true positive rate, defined as the proportion of queries for which at least one relevant passage from the ground-truth article is retrieved among the top-$k$ results (with $k=5$ in all experiments). For SPAR, retrieval and querying are decoupled into two commands: (1) file retrieval, which returns relevant documents, and (2) normal querying, which answers user questions using the workspace database. To ensure fairness, we apply the same set of questions to both systems, but strip away the multiple-choice options when performing file retrieval for SPAR. Sample questions are provided in Appendix~\ref{sec:app_example}.  

\textbf{Answer accuracy.}  
Answer accuracy is assessed by the true positive rate of the LLM agent’s responses to the full multiple-choice questions, i.e., the proportion of cases where the correct answer is selected. Both SPAR and the baseline system use the same agent—Qwen2.5-VL 3B~\cite{qwen_vl}—for answer generation, ensuring that differences arise solely from retrieval design.

\subsection{Results and Interpretation}
\begin{table}[th]
\centering
\caption{Preliminary quantitative results of two RAG systems on the toy biomedical corpus.}
\label{tab:result}
\begin{tabular}{@{}lll@{}}
\toprule
                        & Ordinary RAG & SPAR (Ours) \\ \midrule
Retrieval Accuracy      & 80.3\%       & 89.5\%      \\
Average Retrieval Time  & 0.039s       & 0.015s      \\
Answer Accuracy         & 65.1\%       & 68.1\%      \\ \bottomrule
\end{tabular}
\end{table}

Table~\ref{tab:result} reports the quantitative comparison between the ordinary RAG baseline and SPAR. Overall, SPAR demonstrates consistent improvements in retrieval accuracy, answer accuracy, and efficiency while using the same LLM agent.  

Specifically, SPAR achieves a $9.2\%$ absolute improvement in retrieval accuracy. This gain highlights the benefit of hierarchical routing and metadata-aware filtering, which narrow the candidate search space and improve relevance. Importantly, to ensure fairness, we did not tailor the queries or prompts specifically for file retrieval. With optimized prompting strategies, we expect the retrieval gap could widen further in favor of SPAR.  

In addition, the average retrieval time in SPAR is reduced by more than half compared to ordinary RAG ($0.015$ second vs.\ $0.039$ second), reflecting the efficiency of constructing smaller, task-specific vector databases rather than operating over a large global index.  

Finally, by providing the LLM with more relevant context, SPAR improves answer accuracy by approximately $3\%$. While the gain is modest, it demonstrates the downstream impact of more precise retrieval and points toward further improvements when combined with optimized retrieval–generation integration.



\section{Discussion}
\label{sec:discussion}

\subsection{Motivation Recap}
As outlined earlier, SPAR prioritizes resource efficiency by avoiding global vector indexes when working with large legacy file-based system and instead relying on on-demand, task-scoped workspaces. This design reduces persistent storage and maintenance costs while aligning with the nature of enterprise workflows, where retrieval is typically short-term and context-specific. The architecture also emphasizes modularity, transparency, and controllability, all of which are often overlooked in conventional RAG pipelines.

While these benefits are evident, SPAR also introduces limitations when applied outside its intended scope. We discuss these trade-offs below.

\subsection{Trade-offs of On-Demand Workspace Design}
Because vector indexes are constructed on demand, each workspace must preprocess and embed its candidate file set whenever retrieval is initiated. For small or infrequent tasks this overhead is acceptable, but in larger sessions or environments where similar workspaces are repeatedly created, the setup cost becomes more pronounced.

To reduce redundant computation, we cache normalized and embedded file representations at the file level. This prevents re-embedding for previously processed files, but filtering and index construction must still be repeated for each workspace. In high-throughput or multi-user settings, this design may result in avoidable overhead. 

Another challenge is storage redundancy. Files reused across different sessions may be embedded multiple times into separate workspace-local indexes. Although these indexes are ephemeral—archived or discarded once a workspace is closed—the transient duplication still increases memory and disk footprint when many workspaces are active simultaneously. Without further optimizations such as deduplication or shared indexing, storage costs can scale linearly with the number of concurrent workspaces. 

In practice, we expect these issues to emerge primarily under atypical usage patterns (e.g., excessive or overlapping workspace creation). For the scoped, project-oriented tasks common in enterprise environments, SPAR is likely to operate efficiently. Nonetheless, future improvements such as incremental index construction or cross-workspace sharing mechanisms could further mitigate redundancy.

\subsection{Dependency on Metadata Quality}
SPAR's file filtering logic relies on the structured metadata (e.g., timestamps, enterprise-defined tags) as a prerequisite. In practice, however, metadata may be incomplete, inconsistently applied, or altogether absent. Enterprise-defined tags, in particular, are often curated manually, which can make them noisy, sparse, or unavailable in domains where ongoing maintenance is resource-intensive.

To alleviate this burden, we proposed LLM-assisted file tagging and hierarchy construction. While effective in scaling tag assignment, this approach introduces its own trade-offs. First, consistency: similar files may be assigned slightly different tags, reducing retrieval reliability. This can be addressed through normalization rules, clustering-based reconciliation, or periodic audits with human oversight. Second, interpretability: automatically generated hierarchies may suffer from over-grouping (merging distinct concepts) or mis-grouping (placing tags under inappropriate parents). These issues can be mitigated with lightweight expert supervision, constraint-based clustering, or iterative refinement. Despite these risks, hybrid strategies—enterprise-defined leaf tags combined with LLM-augmented grouping—offer a pragmatic balance between scalability and semantic control.

When metadata is missing or unreliable, recall becomes the primary concern. Relevant documents may be excluded too early in the retrieval process, preventing them from reaching the embedding stage. This limitation is especially acute in legacy systems with inconsistent archival practices. SPAR partially mitigates this issue by exposing the tag and metadata predicates used in retrieval, giving users visibility into the filtering logic and the ability to relax overly strict constraints. However, the root cause lies in organizational data hygiene: robust metadata management and standardization are prerequisites for maximizing SPAR’s effectiveness.

\subsection{Summary of Limitations and Future Directions}
In summary, SPAR trades persistent global indexing for on-demand workspace construction, achieving efficiency and controllability at the cost of repeated setup overhead and potential redundancy under heavy use. Likewise, its reliance on metadata offers interpretability and precision but inherits the weaknesses of incomplete or inconsistent metadata ecosystems. These trade-offs highlight opportunities for future research, such as incremental or shared indexing across workspaces, adaptive caching policies, and more reliable metadata generation pipelines. Addressing these challenges would further strengthen SPAR’s practicality for deployment in diverse enterprise environments.

\bibliographystyle{unsrt}  
\bibliography{references}  

\newpage
\appendix
\section{Appendix}
\label{sec:appendix}

\subsection{Implementation details of SPAR in Biomedical Corpus application.}
\label{sec:appendix_implementation}
\subsubsection*{System Construction}
We implement both the ordinary RAG baseline and the SPAR prototype by extending the codebase of the open-source RAG framework \textit{AnythingLLM}~\cite{anythingllm}.  

For the ordinary RAG pipeline, the system is backed by a Pinecone vector database~\cite{pinecone}, which stores dense embeddings for the entire corpus. Querying and response generation are handled by an open-source LLM chatbot agent integrated through the Ollama framework~\cite{ollama}. This setup reflects the standard retrieve-and-generate paradigm widely adopted in practice.  

For SPAR, we replace the global vector index with a \textbf{Metadata Index} implemented in PostgreSQL~\cite{postgresql}, following the relational schema described in Table~\ref{tab:scheme}. To support tag extraction and comparison during user retrieval commands, we additionally maintain a lightweight Pinecone vector database containing embeddings of the tag vocabulary. This auxiliary index enables semantic similarity checks over tags while the Metadata Index governs structural filtering and hierarchy management. Together, these components realize SPAR’s session-based retrieval logic.

\subsubsection*{Prompts used to query LLM Agent}
To generate our set of multiple choice questions, the following prompt is fed into Qwen2.5-VL 7B along with the corresponding text articles:
\begin{promptbox}
[ARTICLE\_METADATA] \\
Title: {TITLE} \\
PMID: {PMID} \\
MeSH tags: {MESH\_TAGS\_COMMA\_SEPARATED} \\

[ARTICLE\_MAIN\_TEXT] \\
{PLAIN\_TEXT\_MAIN\_BODY} \\

[TASK] \\
Create ONE grounded 4-option MCQ as JSON following the schema: 
\begin{verbatim}
{
  "pmid": "<PMID>",
  "mesh_tags_used": ["<MeSH 1>", "<MeSH 2>", "..."],   // choose 1–3 most relevant
  "question": "<clear, self-contained stem>",
  "options": { "A": "...", "B": "...", "C": "...", "D": "..." },
  "correct_option": "A|B|C|D",
}
\end{verbatim}
Target any of these facet (pick ONE that the text supports best): \\
- precise finding (Results), \\
- mechanism/pathway explanation, \\
- methodology detail (e.g., study design, sample, assay), \\
- clinical implication or limitation/assumption. \\

Constraints \& style: \\
- The stem must be answerable from the provided text alone. \\
- Options A–D must be plausible and mutually exclusive; only one is fully supported by the text. \\
- If the text supports multiple options, revise distractors to be close but definitively incorrect (alter condition, scope, comparator, unit, cohort, or time frame). \\
- Avoid ``not/except'' stems unless absolutely necessary.

If a unique correct answer cannot be guaranteed, output:
\begin{verbatim}
{
  "skip": true,
  "reason": "unanswerable or multiple correct"
}    
\end{verbatim}

[OUTPUT] \\
JSON only. No commentary.
\end{promptbox}

\subsection{Examples of multiple choice questions}
\label{sec:app_example}
In this section, we present some MCQs generated by Qwen2.5-VL 7B for the toy experiment.

For article~\cite{example_1} that studies residential greenness and allergic diseases, here is the generated question:
\begin{promptbox}
    \begin{verbatim}
    Which satellite product and resolution did the study use to compute NDVI for residential 
    greenness?
    
    A. Landsat-8 OLI; 30 m, monthly, UTM projection
    B. Terra MODIS MOD13A3 (v6.1); 1 km, monthly, sinusoidal projection
    C. Sentinel-2 MSI; 10 m, weekly, WGS-84 projection
    D. VIIRS; 500 m, daily, Lambert conformal projection
    
    Correct answer: B
    \end{verbatim}
\end{promptbox}
Similarly, for ~\cite{example_2}:
\begin{promptbox}
    \begin{verbatim}
    In antibiotic-treated mice colonized with Streptococcus mutans, which observation best 
    supports that mTORC1 activation mediates the PD-like pathology?

    A. Rapamycin lowered plasma and brain imidazole propionate (ImP) to baseline but did not 
    change neuronal pathology.
    B. Rapamycin left pS6/4E-BP1 phosphorylation unchanged yet improved motor function.
    C. Rapamycin suppressed mTORC1 signaling in TH+ neurons and rescued neurodegeneration and 
    motor deficits, even though ImP levels remained elevated.
    D. Pirfenidone reduced dopaminergic neurodegeneration in vivo without affecting ImP levels.
    
    Correct answer: C
    \end{verbatim}
\end{promptbox}

For ~\cite{example_3}:
\begin{promptbox}
    \begin{verbatim}
    During the study period, when did Australia’s HCV DAA dispensings peak, and what was the 
    approximate count?
    
    A. March 2016; ~11,400 dispensings
    B. June 2016; ~20,200 dispensings
    C. July 2018; ~10,000 dispensings
    D. January 2021; ~4,000 dispensings
    
    Correct answer: B
    \end{verbatim}
\end{promptbox}

\end{document}